\documentclass[fleqn,10pt]{wlscirep}
\usepackage[utf8]{inputenc}
\usepackage[T1]{fontenc}
\usepackage{graphicx}
\usepackage{caption}
\usepackage{subcaption}

\definecolor{forestgreen(web)}{rgb}{0.13, 0.55, 0.13}

\title{Food Pairing Unveiled: Exploring Recipe Creation Dynamics through  Recommender Systems}

\author[1,2,*,+]{Giovanni Palermo}
\author[3,+]{Claudio Caprioli}
\author[2,+]{Giambattista Albora}

\affil[1]{Physics Department, University "Sapienza", 00185 Rome (Italy)}
\affil[2]{Enrico Fermi Research Center, 00184 Rome (Italy)}
\affil[3]{Department of Physics and Astronomy ”Ettore Majorana”, University of Catania}

\affil[*]{giovanni.palermo@cref.it}

\affil[+]{these authors contributed equally to this work}

\begin{abstract}
In the early 2000s, renowned chef Heston Blumenthal formulated his "food pairing" hypothesis, positing that if foods share many flavor compounds, then they tend to taste good when eaten together.
In 2011, \emph{Ahn et al.} conducted a study using a dataset of recipes, ingredients, and flavor compounds, finding that, in Western cuisine, ingredients in recipes often share more flavor compounds than expected by chance, indicating a natural tendency towards food pairing. \\Building upon Ahn's research, our work applies state-of-the-art collaborative filtering techniques to the dataset, providing a tool that can recommend new foods to add in recipes, retrieve missing ingredients and advise against certain combinations.\\We create our recommender in two ways, by taking into account ingredients appearances in recipes or shared flavor compounds between foods.
While our analysis confirms the existence of food pairing, the recipe-based recommender performs significantly better than the flavor-based one, leading to the conclusion that food pairing is just one of the principles to take into account when creating recipes. Furthermore, and more interestingly, we find that food pairing in data is mostly due to trivial couplings of very similar ingredients, leading to a reconsideration of its current role in recipes,
from being an already existing feature to a key to open up new scenarios in gastronomy. Our flavor-based recommender can thus leverage this novel concept and provide a new tool to lead culinary innovation.
\end{abstract}
\begin{document}

\flushbottom
\maketitle

\thispagestyle{empty}

\section*{Introduction}
The past two decades have witnessed what is usually referred to as the Big Data revolution: the rapid increase of collected data, the accelerating data storage and computing capacity and the breakthroughs in machine learning have given rise to entire new research areas in disparate scientific fields, collecting a series of astonishing results and solving some of the most challenging scientific problems formulated in the second half of the 20th century~\cite{hey2009the}. While the major changes have mainly affected fields as biology and linguistics~\cite{Khurana2023}, gastronomy, a field rarely touched by scientific enquiry in the past, has recently sparked the interest of many scientists~\cite{Ahnert2013}, leading to the birth of computational gastronomy~\cite{goel2022computational}.\\
Among many of the possible research questions, a preeminent aspect that interests researchers studying this new field, is understanding the reason behind food palatability~\cite{lawless2010sensory}, the degree of pleasure and satisfaction derived from eating a particular food, influenced by a combination of its sensory attributes. These attributes include taste, aroma, texture, appearance, temperature, and overall flavor. 
This desire led to the emergence of hypotheses such as food pairing, a well-known gastronomical conjecture introduced by Chef Heston Blumenthal~\cite{blumenthal2008big}, positing that ingredients sharing flavor compounds tend to be more compatible with each other. 
A task of computational gastronomy is to provide evidence that supports or rejects these hypotheses. In this context, Ahn et al. were pioneers in using network methods to try to provide evidence in favor of food pairing, stating that it is verified only for North American cuisine, while the Asian one (Chinese, Japanese, Korean) behave the opposite.
This paved the way for subsequent studies, such as those by Jain et al.~\cite{Jain2015SpicesFT}, who focused on Indian cuisine, Varshney et al.~\cite{Varshney2013FlavorPI}, who explored medieval cuisine, and Simas et al.~\cite{Simas2017}, who also investigated food bridging, a hypothesis that involves the idea that a third ingredient can serve as a bridge to enhance the compatibility between two other ingredients that do not share flavor compounds, thereby enriching the gastronomical experience.\\
Validating or negating these hypothesis is key to understanding the mechanism behind the formation of recipes and the perception of taste. 
The addition or replacement of ingredients in recipes is an increasingly popular demand that, in recent years, has intertwined with more compelling problems. Being able to replace ingredients in recipes could on one hand allow to make food more sustainable by responding to environmental issues, and on the other hand meet health needs, such as improving nutrient content of food or adapting recipes to food intolerance. \\
In recent years, fueled by vast amount of collected data and recently developed data-mining methods,
recommender systems have been used to tackle these pressing demands, leading to the development of food recommendation, a field that has rapidly come to be a research area of its own~\cite{BONDEVIK2024122166, pellegrini2021exploiting}.\\
Our work starts by observing that food pairing hypothesis can be naturally associated to item-item collaborative filtering, a particular kind of recommender system~\cite{sarwar2001item} that, starting from a bipartite network that connects item to users according to their preferences, works by recommending an item I to user U if U is connected to items similar to I. In food recommendation, recipes can be regarded as users and ingredients as items: an ingredient I is recommended to recipe R if R contains ingredients that are very similar to I. In this context, food pairing postulates that the similarity between pairs of ingredients is positively correlated to the number of flavor compounds they have in common.\\ 
Leveraging on this, starting from the same recipe data used in the work of Ahn et al.~\cite{ahn2011flavor},  we compare different recommenders using ingredient-ingredient similarity based on flavor compounds. We contrast these findings with a system that bases its recommendations on recipe co-occurrence data. Our analysis reveals that the recipe-based approach generally surpasses the flavor-based system in terms of accuracy for ingredient recommendations, underscoring the limited role of food pairing in recipe creation. 
Furthermore, within the recipe-based case, we compared the conventional similarity-based systems with LightGCN, a graph convolutional network. As previously observed in the literature~\cite{Liu_2022}, LightGCN tends to yield more accurate recommendations. However, this increase in accuracy comes at the cost of reduced interpretability due to the opaque, 'black box' nature of machine learning algorithms.
Recommenders based on recipe co-occurrence prove quite successful due to the lack of assumptions on ingredient association mechanisms, extracting association rules thorugh purely statistical co-occurrence properties. On the contrary, the flavor-based recommender system demonstrates efficacy primarily within North American culinary contexts, corroborating Ahn et al.'s conclusions regarding the regional specificity of food pairing. Further scrutiny into the nature of ingredient compatibility highlighted that the apparent success of food pairing often results from simplistic matches between highly similar ingredients (for example, 'bell pepper' and 'green bell pepper') which share identical flavor profiles. This observation prompts a reevaluation of the underlying mechanisms attributed to food pairing in culinary science.

\section*{Results}

The backbone of our study is Ahn's dataset of recipes, ingredients and flavor compounds. Each of these three entities can be seen as a layer of a tripartite network (see Methods), in which each recipe is linked to the ingredients it is made up of and each ingredient to the flavor compounds it contains. In this section, we will show our findings when using this dataset to set up a collaborative filtering and using it to investigate the food pairing hypothesis.\\
\begin{figure}[h]
        \centering
        \includegraphics[width=\linewidth]{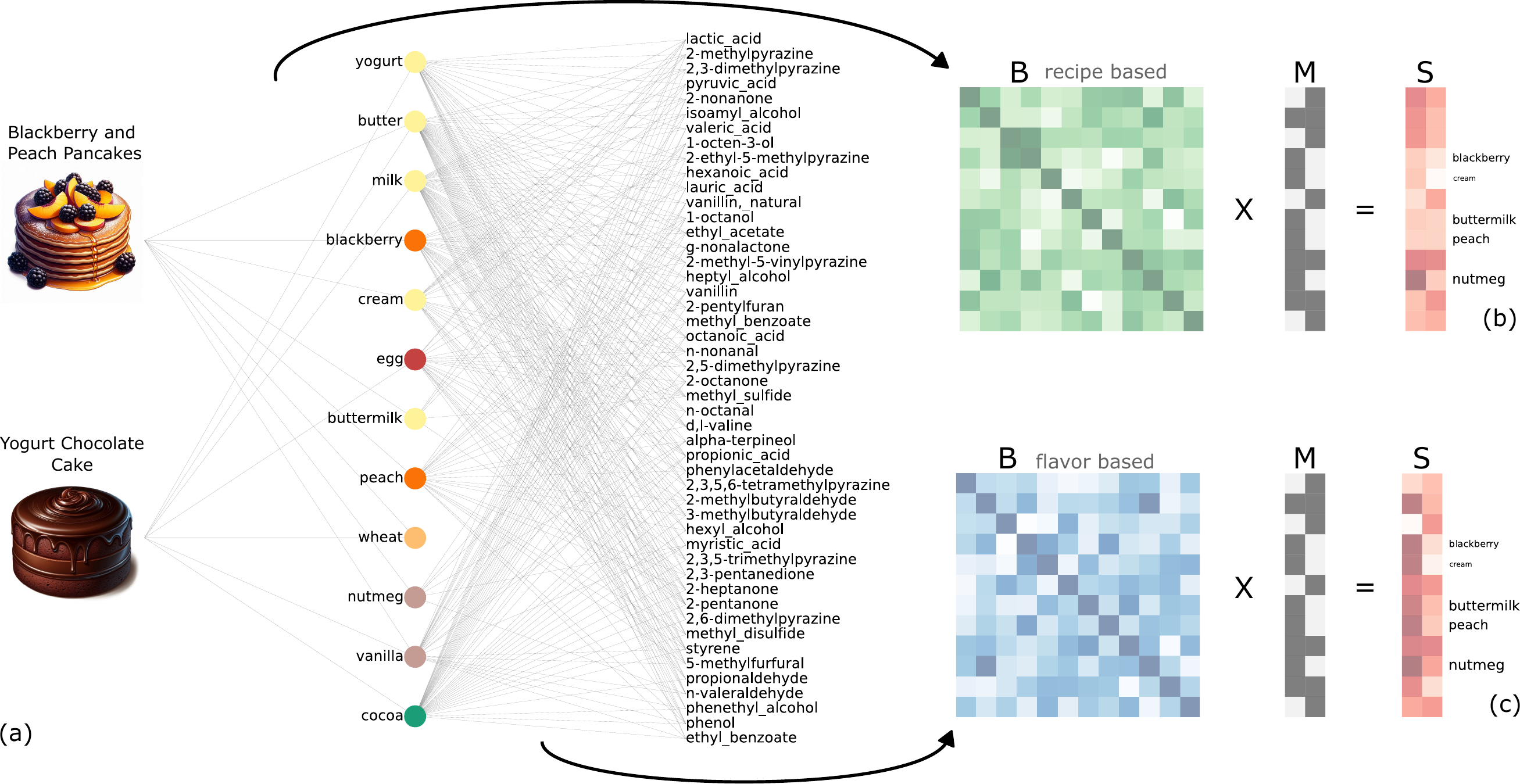}
        
    \caption{Schematic representation of dataset and algorithm: (a) each recipe is connected to the ingredients it is made up of, and each ingredient to its flavor compounds. Collaborative filtering can work in two different ways. (b) exploits the cooccurrences of ingredients in recipes to recommend a new one for the chosen recipe; e.g. it suggests to add blackberries to a yogurt chocolate cake, as many ingredients in this cake are often used together with blackberries. (c) recommends an ingredient according to the flavor compounds it shares with the other ones in the recipe; e.g. buttermilk is recommended since it shares many flavor compounds with yogurt and butter, that are already in the recipe.}
    \label{recipe_based_comparison}
\end{figure}
\subsection*{Recipes-Ingredients based recommender}
In the first place we built an item-item collaborative filtering to exploit the first two layers (recipes-ingredients) of the tripartite network to make recommendations of ingredients.  We computed the similarity between ingredients according to their cooccurrences in recipes, i.e. two ingredients are similar if they appear in the same recipes many times. Thus, an ingredient is as strongly recommended for a recipe as many similar ingredients the recipe contains. By doing so, we got a quick and efficient way to obtain recommendations by taking into account pairwise interactions between ingredients. Additionally, to quantify the limitations of accounting for only pairs of ingredients in the computation of similarity, disregarding the context in which they are used, we used LightGCN, a graph convolutional network that overcomes this limitation thanks to a global view of the network structure not limited to simple cooccurrences\footnote{See Methods for a detailed description of how to compute the similarity.}.\\
To assess the performance of our recommender, we tested its ability of retrieving missing ingredients from recipes. Thus, we first removed one random ingredient from the $10\%$ of recipes, computed the similarity between ingredients and checked, for each recipe, whether the missing ingredient was among the top recommendations.

\begin{figure}[h]
        \centering
        \includegraphics[width=\linewidth]{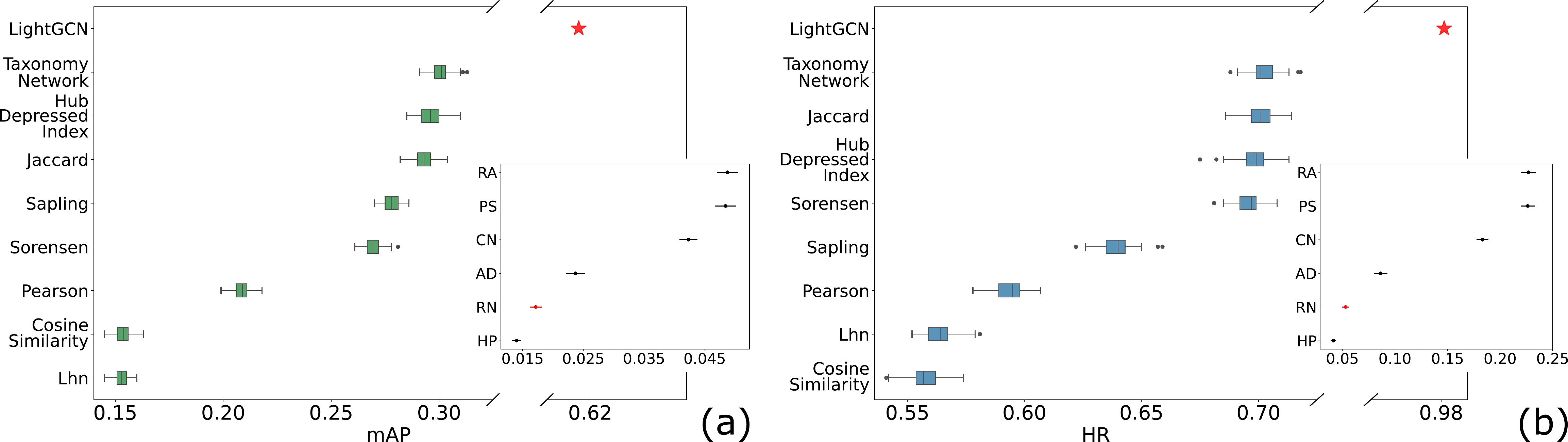}

    \caption{(a) Mean Average Precision (mAP) and (b) Hit Ratio at 20 (HR@20) for all the methods used, averaged over 100 runs with an error bar equal to one standard deviation.  The red point accounts for the performance of random recommendations.}
            \label{recipe_based_comparison}
\end{figure}

Figure \ref{recipe_based_comparison} shows a comparison between the performance of all the methods used to get the recommendations, with the error bars equal to one standard deviation over 100 runs. We included the performance of random recommendations in red on the graph.\\
Our results can be summarized as follows:
\begin{itemize}
    \item All methods perform significantly better than a random model.
    \item LightGCN outperforms all the other metrics, showing the relevance of the context in which ingredients are used in recipes.
    \item The other methods are split in two clusters; the highest score ones (LHN and above) are the ones that take into account the degree of both ingredients when computing the similarity, in order to avoid to over-recommend ingredients only due to their vast presence in the dataset. Instead, the other methods are affected by this issue and perform much worse.
\end{itemize}
LightGCN is the only method for which we have only one run, due to the high computational time required for each training of the model. However, its performance is significantly higher than the other ones, allowing us to confirm the importance of higher order interaction beyond pairwise to make recommendation.
\\\\
These results provide a reliable recommender for adding or substituting ingredients in recipes, making it easy to satisfy various constraints, such as getting a healthier recipe, one that uses more sustainable foods or simply replace an undesired ingredient (due to allergies or personal taste).

\subsection*{Ingredients-Flavor compounds based recommender}
Since the dataset also includes the flavor compounds, the next step was to exploit this information to base the recommendation of an ingredient on the shared flavor compounds, in line with the food pairing hypothesis.\\
First, we checked if our findings were compatible with the ones in \cite{ahn2011flavor}, namely if existing recipes already share more flavor compounds than they would at random (at least for the North American cuisine). For this purpose, we used the ingredient-flavor compounds layers to compute the similarity between ingredients, then we made recommendations according to this new similarity.

\begin{figure}[h]
        \centering
        \includegraphics[width=\linewidth]{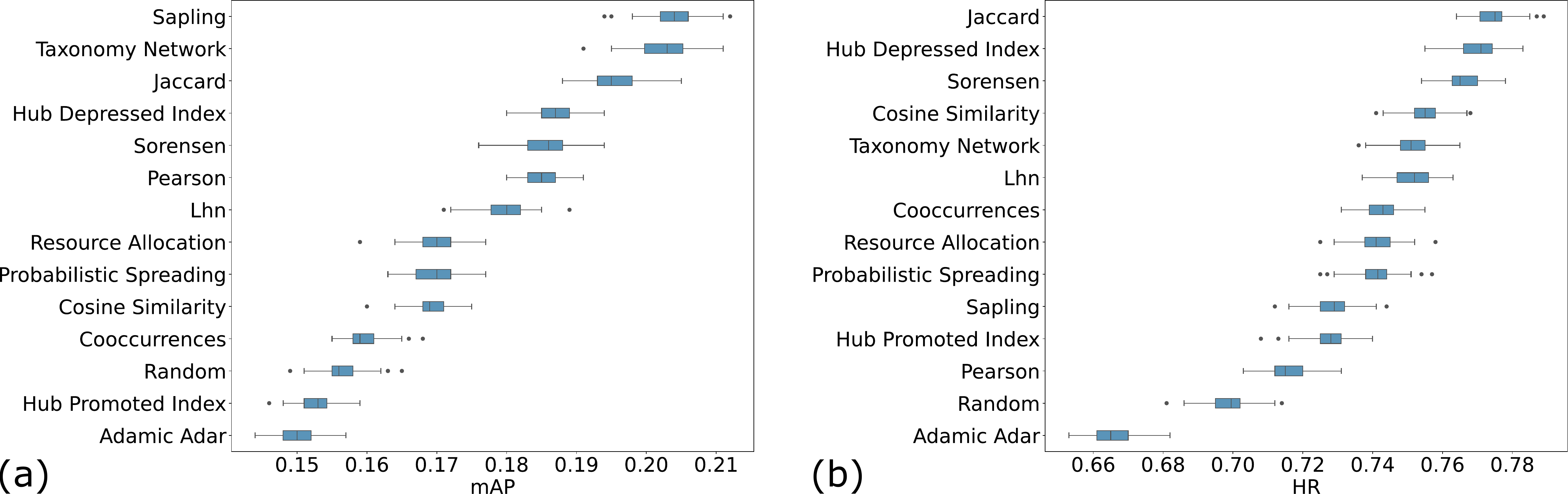}
        
    \caption{Same plots as in \ref{recipe_based_comparison} but computed with flavor compounds similarity.}
    \label{compound_based_comparison}
\end{figure}

In this case, we restricted the recommended ingredients to the ones in the category of the one to be retrieved, otherwise, each recipe would have been matched with similar ingredients already contained in it. We provide an example for the seek of clarity: suppose we have a recipe for making custard cream, which is made up of 'milk', 'cream', 'butter' and 'yolk'. We remove 'yolk' from the recipe and want to retrieve it with the recommender, which computes the similarity between all the possible ingredients and the other ones in the recipe. Since these last are all dairies and the similarity is computed according to the shared flavor compounds, the top recommendations will all be very similar to 'milk', 'cream' and 'butter', such as 'buttermilk' or 'mozzarella'. These would be of course trivial matches that share many flavor compounds with the other ingredients just because they belong to the same category, therefore we search only among the category 'animal product' (which contains 'yolk'); if the missing ingredient is retrieved, it will mostly be thanks to food pairing.\\
As shown in Figure~\ref{compound_based_comparison}, most of the metrics used perform better than a random model, confirming food pairing already exists in recipes. Nevertheless, it is clear that this recommender based on flavor compounds performs worse than the one based on recipes. Indeed, although food pairing may be true and foods sharing many flavor compounds may actually be good eaten together, there are other factors influencing taste (citazione), like temperature or texture. Therefore, the cooccurrences of ingredients in recipes capture more of these factors by simply considering when foods are used together.\\
Overall, these results confirm the best methods to be the ones of the recipe-based recommender. Yet in this case we did not use LightGCN, as it would not be straightly connected to food pairing. Indeed, while the item-item collaborative filtering takes into account the cooccurrences of flavor compounds, LightGCN may find more complex correlations between recipes and flavor compounds. For example, it might learn to match the flavor compounds of butter and eggs for the simple reason that they are widely used together in recipes, and not because they share many flavor compounds.

\subsubsection*{Food pairing across the world}
We then repeated the analysis of the previous subsection, restricting the recipes to specific regions of the world \footnote{We use the same mapping region-country as in \cite{ahn2011flavor}}. In this way, we aimed to retrieve Ahn's results for food pairing in each region.

\begin{figure}[h]
    \centering
    \includegraphics[width=\linewidth]{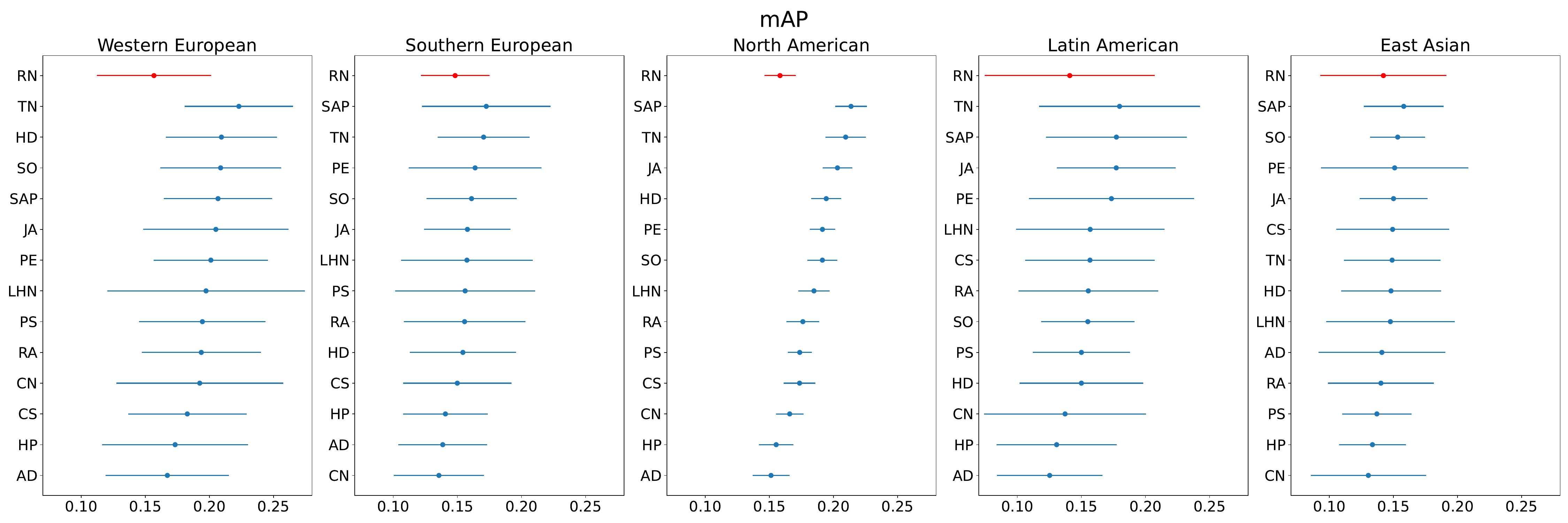}
    \caption{mAP for the recommenders built for each regional cuisine, reported with one standard deviation error bars, along with random recommender in red.}
    \label{compounds_scores_region}
\end{figure}

In Figure \ref{compounds_scores_region} we show the flavor compound-based recommender performs significantly better than a random model only in the case of North American cuisine\footnote{Ahn's result for East Asian cuisine sharing less flavor compounds than a random model is compatible with our results within the error.}. This result confirms Ahn's findings and, more importantly, corroborates the use of collaborative filtering to test the food pairing hypothesis.

\subsection*{How Ingredients Shape Recommendations}
The results presented so far are all in terms of global metrics accounting for average performances (mAP and HR@20). Here we want to investigate how our recommenders work at a finer level, looking at the role of single ingredients when computing the scores of the recommendations. In this way, we can highlight if food pairing concerns specific pairs of ingredients or if it is a widespread and even tendency in recipes. To do so, we inspect the similarity between the missing ingredient and the other ones in the recipe, and then compute the Z-score of the similarities. The Z-score is the number of standard deviations by which the value of a data point (in our case one similarity between foods) is above or below the mean value of all the observations (all the other similarities with the ingredients in the recipe).

\begin{figure}[h]
    \begin{subfigure}[b]{0.49\textwidth}
        \centering
        \includegraphics[width=\linewidth]{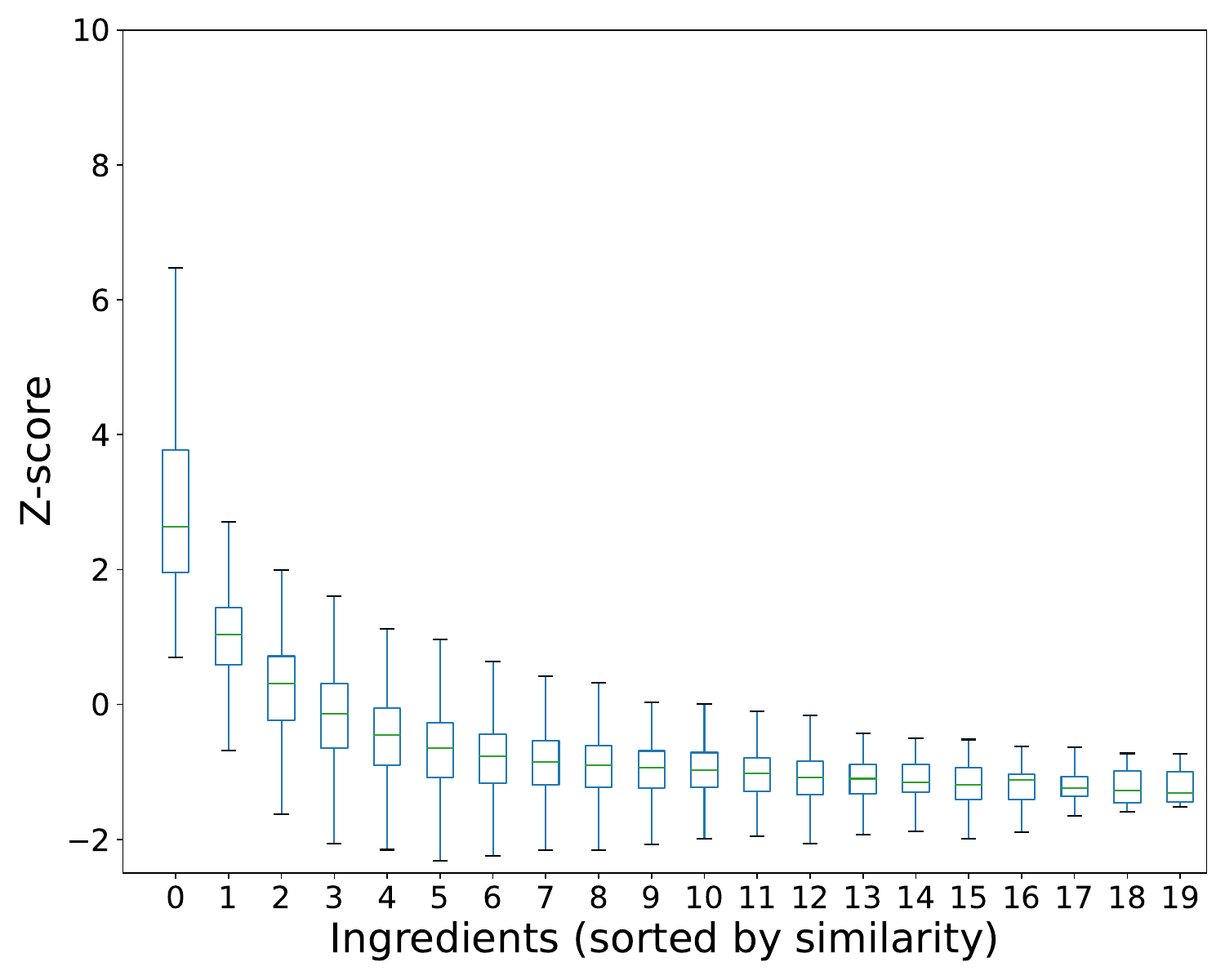}
        \subcaption{}
        \label{metrics}
    \end{subfigure}
    \begin{subfigure}[b]{0.49\textwidth}
        \centering
        \includegraphics[width=\linewidth]{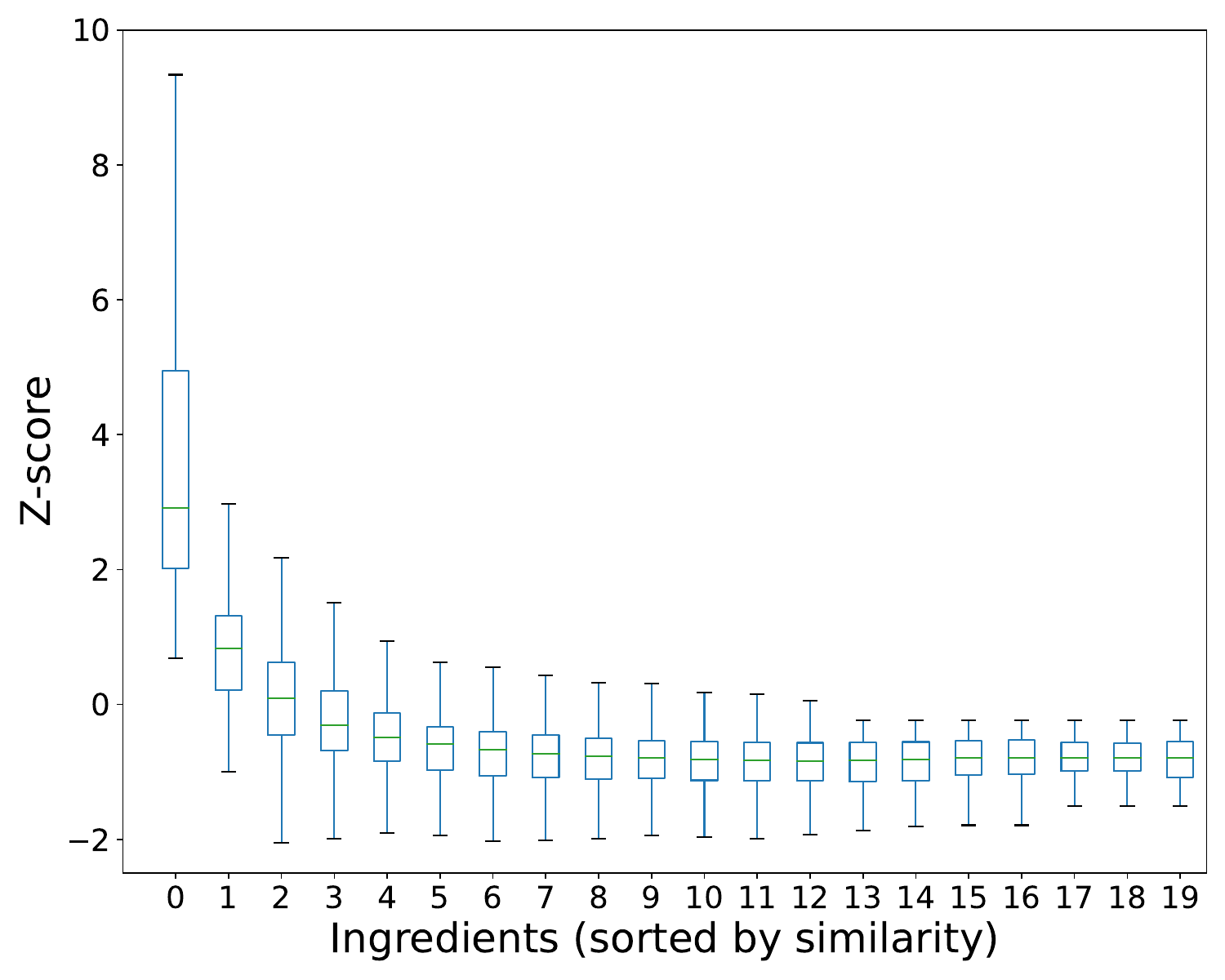}
        \subcaption{}
        \label{metrics_comp}
    \end{subfigure}
    \caption{Boxplot of the Z-Scores between the missing ingredient and the other ingredients in the recipe (first 20 ingredients only, recipes contain no more than 32 ingredients each), ranked by similarity, for the recipe-based (a) and the flavor compound-based (b) recommendations.}
    \label{zscores}
\end{figure}

The Z-scores in Figure \ref{zscores} provide a crucial insight to understand the nature of matches between foods. When the missing ingredient is successfully retrieved by the recommender, it is due to the ingredient in the recipe with the highest similarity with it, as the Z-score of this similarity is significantly high. This effect becomes extreme for the flavor compound-based recommendations, in which the Z-score of the most similar ingredient to the missing one is almost above $2$ in the $73\%$ of cases. Indeed, when points are normally distributed, only $4.5\%$ of the observations fall farther than $2$ standard deviations from the mean.\\
This finding is a strong indication to inspect what happens at the single ingredient level. Our analysis reveals the food pairing to be a trivial consequence of the addition of many similar ingredients into recipes. We show it by plotting how many times an ingredient is retrieved and what is the most similar ingredient in the recipe (and is thus responsible for the success of the algorithm, as proved by Figure \ref{zscores}). To do so, we selected four distinctive families of ingredients with high presence in the recipes, i.e. 'amarylidacee' (the family of onions, garlic and similar), 'peppers', 'cheese' and 'citrus fruits'; hence, for instance, all different kinds of cheese are grouped into a single word 'cheese'.

\begin{figure}[h]
    \begin{subfigure}[b]{0.49\textwidth}
        \centering
        \includegraphics[width=\linewidth]{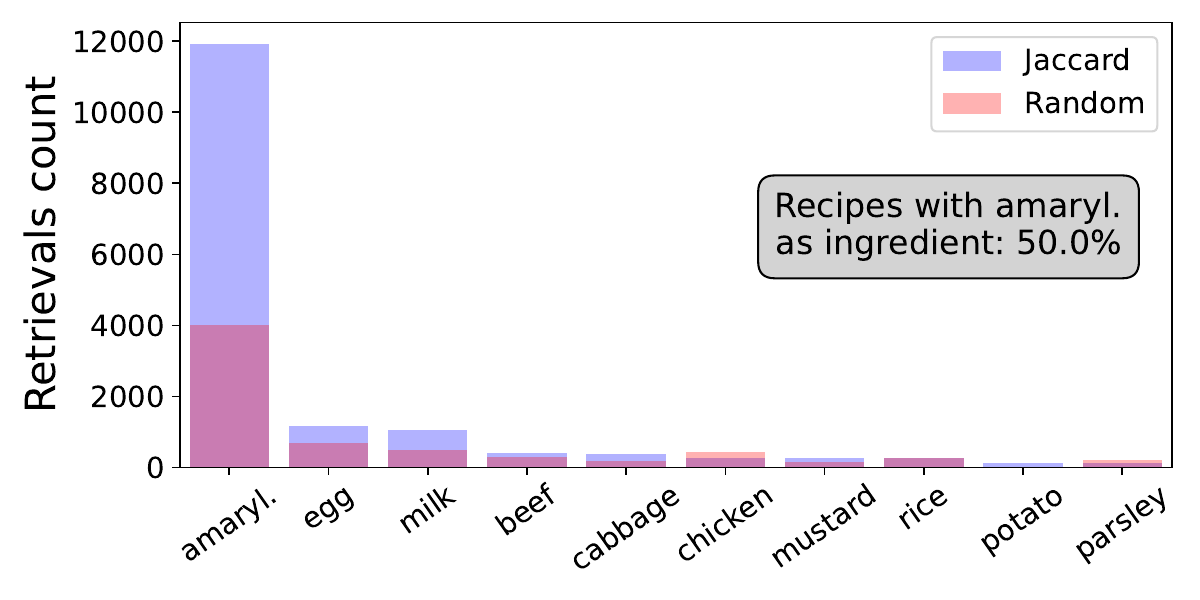}
        \subcaption{}
        \label{amary}
    \end{subfigure}
    \begin{subfigure}[b]{0.49\textwidth}
        \centering
        \includegraphics[width=\linewidth]{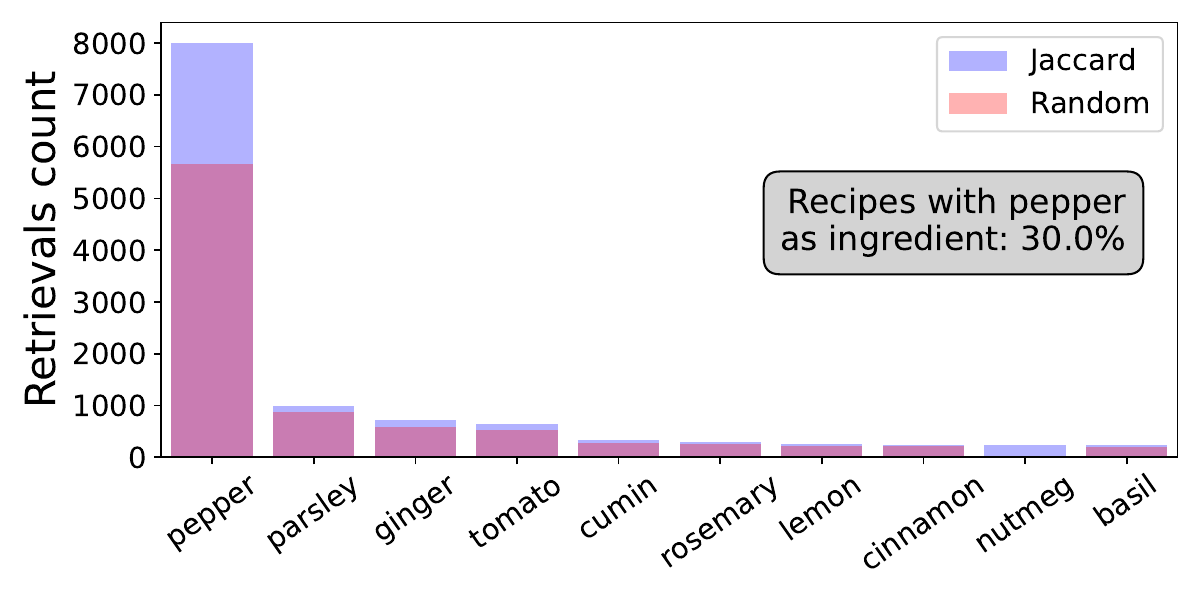}
        \subcaption{}
        \label{pepper}
    \end{subfigure}
        \begin{subfigure}[b]{0.49\textwidth}
        \centering
        \includegraphics[width=\linewidth]{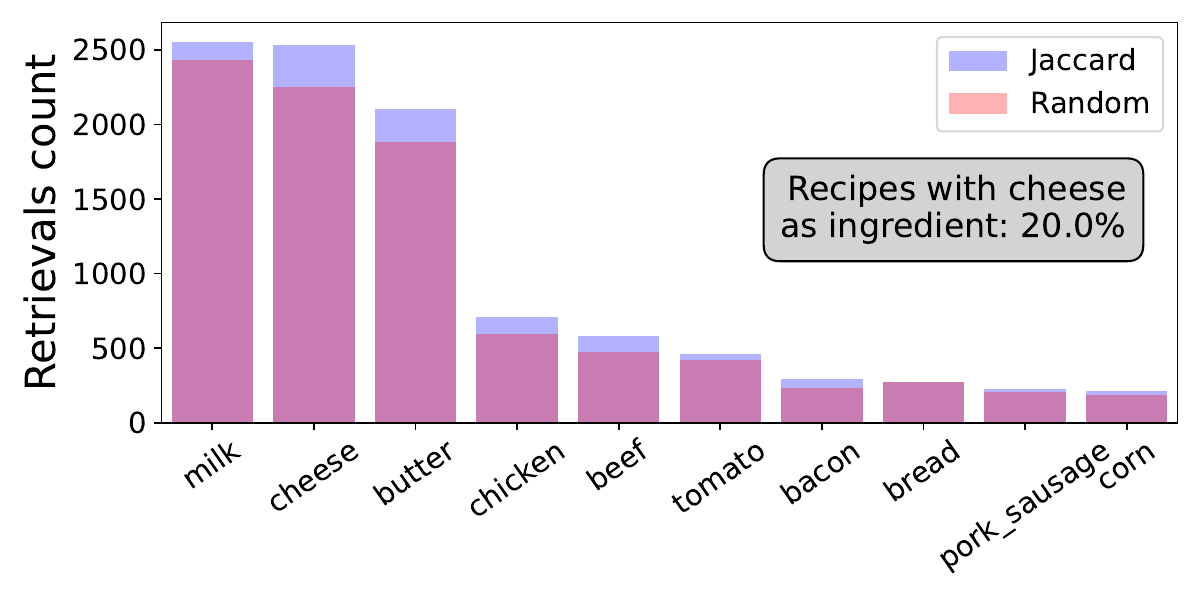}
        \subcaption{}
        \label{cheese}
    \end{subfigure}
    \begin{subfigure}[b]{0.49\textwidth}
        \centering
        \includegraphics[width=\linewidth]{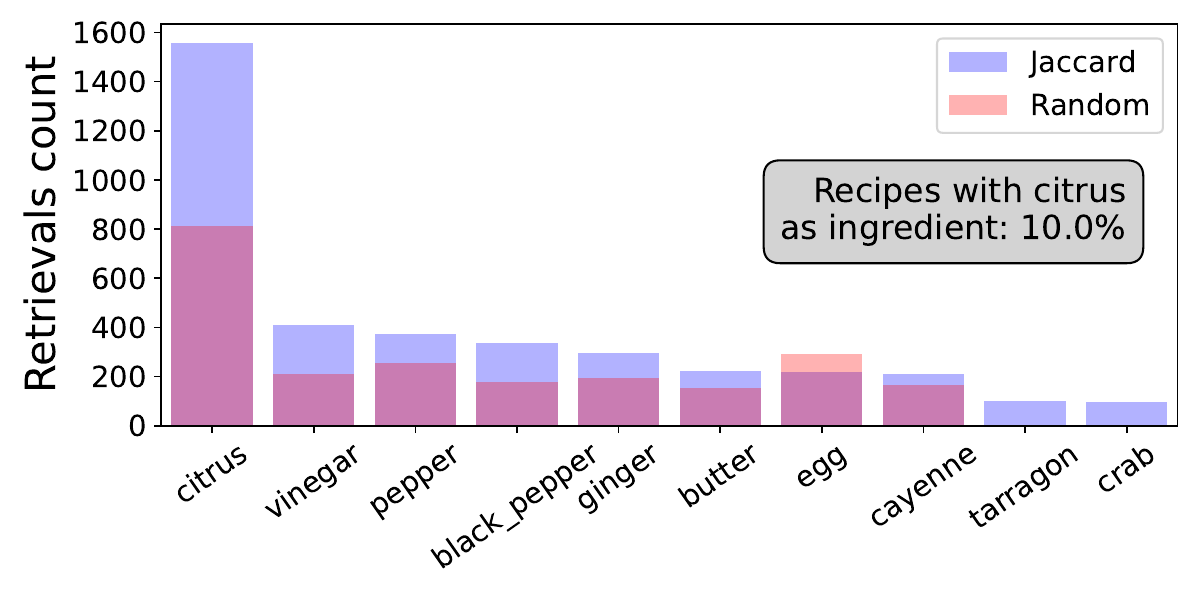}3
        \subcaption{}
        \label{citrus}
    \end{subfigure}
    \caption{Counts of retrieval times for 'amaryllidace' (a), 'pepper' (b), 'cheese' (c) and 'citrus' families (d) as a function of the most similar ingredient in recipes. In this case, we cancel  one ingredient of the chosen family in every recipe where they appear and then check if it is in the top $20$ flavor compound-based recommendations computed with Jaccard (in blue), while in red we guess ingredients randomly. The difference in height between the bars shows how better (or worse) the recommender performs compared to the random model for each family/ingredient.}
    \label{families}
\end{figure}

In Figure \ref{families} we plotted both the results for Jaccard similarity (the metric with the best performance) and a random recommendation. Most of the improvement compared to random is given by the presence of ingredients of the same family in the recipe ('onion' is retrieved by similarity with 'shallot', 'bell pepper' by similarity with 'green bell pepper' and so on). For example, there are $28942$ recipes ($50\%$ of the dataset) containing at least one ingredient of the 'amarylidacee' family; they were retrieved in the top $20$ recommendations in the $60\%$ of times, but 7 times out of 10, this retrieval was only due to another ingredient in the same family in the recipe (such as 'onion' and 'shallot'). Additionally, it is almost only that case that contributes to a significant increase in the performance compared to random, since the $97\%$ of the improvement is given by a match with an ingredient of the same family being the most similar one in the recipe. Therefore, most of the food pairing signal is given by these trivial coupling of very similar ingredients and not by actually matching different foods according to flavor compounds. Yet there are still some minor cases in which the model improves compared to random and the ingredient responsible for the retrieval of the missing one is not one of the same family (such as 'vinegar' and 'pepper' in the case of 'citrus' fruits), as well as cases in which the model performs worse, like citrus fruits and eggs, probably due to ubiquity of eggs in recipes rather than food pairing.
\newpage

\section*{Discussion}

In the past years, recommender systems became a ubiquitous feature of many services, especially e-commerce and other web based applications, capable to customize the user experience at many levels.\\
In this work, we created and validated a collaborative filtering on a dataset of recipes, ingredients and flavor compounds. By developing two versions of our recommender, one recipe-based and one flavor-based, we equipped a broad crowd of potential users, from innovation-oriented chefs to cooking newbies, with a powerful tool to make modifications to recipes, add ingredients or replace others. The recommendation can be done by either considering the cooccurences of ingredients in existing recipes or by exploring new potential matches given by food pairing. In this way, many constraints can be satisfied, like getting a healthier recipe or one with more sustainable foods, or simply replacing an undesired ingredient (due to allergy or personal taste).\\
Furthermore, by comparing the performances of the two recommenders, we were able to assess the importance of food pairing in recipes, confirming it is only one of the several factors that interplay to determine palatability.\\\\
The most important finding of our work is the analysis of food pairing at the single ingredient level. We showed how existing recipes already share more flavor compounds than they would at random just because of the tendency to mix similar ingredients together (like 'red pepper' and 'green pepper'). While such a result might look disappointing at a first sight, it actually offers even more chances of innovating cuisine. Indeed, we may not currently create recipes according to food pairing (or at least to a non-trivial version of it), yet Blumenthal's hypothesis may nevertheless true and creating dishes according to shared flavor compounds could actually spark a new direction in gastronomy. As for our knowledge, our flavor-based recommender is currently the best tool for this purpose and can be easily extended to larger datasets or mixed with others\footnote{One could easily average the recipes-based and the flavor-based recommendations.}.\\
Additionally, by inspecting the similarities between pairs of ingredients, we can suggest new potential matches between foods. For example, if we search for the most similar pairs of fruits and vegetables, new and surprising matches arise, like tomatoes and grape or peaches and bell peppers. Of course, there is no way of validating such recommendations beside trying them. That is why further improvements of this work should include collaborations with chefs willing to innovate their cuisine and experiment with science-based gastronomy.\\

\section*{Methods}
\subsection*{Bipartite networks and adopted notation}
A bipartite network is defined as a graph $G=(\Gamma,I,E)$ where $\Gamma$ and $I$ are two sets of nodes (also called layers), and E is the set of all the connections $(\alpha, i)$ between the nodes $\alpha\in \Gamma$ and $i \in I$. Let $|\Gamma|$ be the dimension of the set $\Gamma$ and $|I|$ the dimension of the set $I$: the bipartite network is represented by a $|\Gamma|\times|I|$ binary matrix $R$ called bi-adjacency matrix and defined as:
\begin{equation}
	R_{\alpha i} =
	\begin{cases}
		1 ~~~\text{ if } (\alpha, i) \in E\\[3mm]
		0 ~~~\text{ if } (\alpha, i) \notin E
	\end{cases}	
\end{equation}
So the element of $R$ corresponding to nodes $\alpha$ and $i$ is 1 if there is a link between them.

\subsection*{Dataset}
The foundation of our study is the dataset introduced by Ahn et al. \cite{ahn2011flavor}, comprising a network with three layers (tripartite) corresponding to recipes, ingredients, and flavor compounds. Recipes are connected to the ingredients that constitute them, and ingredients are linked to the aromatic compounds that characterize them. Additionally, the dataset includes information about the country of origin for each recipe and the category of the ingredient (for instance cheese for mozzarella, and vegetable for red pepper).\\
To build this dataset a total of 57692 recipes have been extracted from three culinary repositories. The resulting tripartite network contains 1530 ingredients along
with their 1021 flavor compounds, yet only 383 ingredients are actually used in recipes. \\
We decompose the data into two bipartite networks: the first connects ingredients to recipes with a biadjacency matrix denoted by R, and the second connects ingredients to flavor compounds with a biadjacency matrix denoted by F.
\subsection*{Item-item Collaborative Filtering}

Collaborative Filtering is a technique used in the Recommender System framework to make automatic predictions about the interests of a user by collecting preferences from many users. The idea is that if a person A has the same opinion as a person B on an issue, A is more likely to have B's opinion on a different issue than that of a random person.\\
Various approaches can be adopted to construct a collaborative filtering system. In this study, we employ an item-item collaborative filtering, which measures the similarity between items based on the user-item bipartite network in which links reflect the interests of the users. Over the years, various metrics have been introduced to calculate this similarity, ranging from simple calculations of co-occurrences between two ingredients (how often they appear in the same recipes) to incorporating normalization factors that can be derived from the degree of ingredients or recipes within the bipartite network (how many links they are involved in). In table \ref{tab:similarities} we present the metrics we adopted to measure the similarity between two items $i$ and $j$.\\

\begin{table}[ht]
\centering
\renewcommand{\arraystretch}{1.5}
\begin{tabular}{rcc}
\textbf{method}  & \textbf{formula}   & \textbf{reference}\\
Cooccurrences ($B^{CN}$) & $CO_{ij}$ & \cite{liben2003link}\\
Jaccard ($B^{JA}$) & $\frac{CO{ij}}{u_i+u_j-CO_{ij}}$ & \cite{jaccard1901etude}\\
Adamic/Adar ($B^{AD}$) & $\sum_\alpha\frac{M_{\alpha i}M_{\alpha i}}{log(d_\alpha)}$ & \cite{adamic2003friends}\\
Resource Allocation Index ($B^{RA}$) & $\sum_\alpha\frac{M_{\alpha i}M_{\alpha j}}{d_\alpha}$ & \cite{zhou2009predicting}\\
Cosine Similarity ($B^{CS}$) & $\frac{1}{\sqrt{u_iu_j}}CO{ij}$ & \cite{salton1983introduction}\\
Sorensen index ($B^{SO}$) & $\frac{1}{u_i+u_j}CO{ij}$ & \cite{sorensen1948method}\\
Leicht-Holme-Newman Similarity ($B^{LHN}$) & $\frac{1}{u_iu_j}CO{ij}$ & \cite{leicht2006vertex}\\
Hub depressed index ($B^{HD}$) & $\frac{1}{max(u_i,u_j)}CO{ij}$ & \cite{ravasz2002hierarchical}\\
Hub promoted index ($B^{HP}$) & $\frac{1}{min(u_i,u_j)}CO{ij}$ & \cite{ravasz2002hierarchical}\\
Taxonomy Network ($B^{TN}$) & $\frac{1}{max(u_i,u_j)}\sum_\alpha\frac{M_{\alpha i}M_{\alpha j}}{d_\alpha}$ & \cite{zaccaria2014taxonomy}\\
Probabilistic Spreading ($B^{PS}$) & $\frac{1}{u_j}\sum_\alpha\frac{M_{\alpha i}M_{\alpha j}}{d_\alpha}$ & \cite{zhou2007bipartite}\\
Pearson Correlation Coefficient ($B^{PCC}$) & $\frac{\sum_\lambda(M_{\alpha i}-u_i/N)(M_{\alpha j}- u_j/N)}{\sqrt{\sum_\lambda(M_{\alpha i}-u_i/N)^2}\sqrt{\sum_\lambda(M_{\alpha j}- u_j/N)^2}}$ &  \cite{shardanand1995social}\\
Sapling Similarity ($B^{SAP}$) & $(\frac{CO_{ij}(1-\frac{CO_{ij}}{u_{j}})+(u_{i}-CO_{ij})(1-\frac{u_{i}-CO_{ij}}{N-u_{j}})}{u_{i}(1-\frac{u_{i}}{N})}-1)sign(1-\frac{CO_{ij}N}{u_{i}u_{j}})$ & \cite{albora2023sapling}\\
\end{tabular}
\caption{Metrics of similarity.}
\label{tab:similarities}
\end{table}
Here we indicate users with Greek letters and items with Latin ones. $M$ is the adjacency matrix of the bipartite network, $CO$ is the matrix of the co-occurrences between items ($CO_{ij} = \sum_\alpha M_{\alpha i}M_{\alpha j}$ is the number of co-occurrences between $i$ and $j$), $d$ (diversification) is the degree of users, $u$ (ubiquity) is the degree of items, $N$ is the total number of items in the network, and $sign$ is the sign function (+1 if the argument is greater than 0 and -1 otherwise).
\newline
Once the similarity matrix $B$ is computed,
the recommendation score $S$ for an item $i$ to a user $\alpha$ is computed as:
\begin{equation}
S_{\alpha i}= \frac{\sum_{j}B_{ij}M_{\alpha j}}{\sum_{j}|B_{ij}|}
\label{eq:density}
\end{equation}
Specifically, $S_{\alpha i}$ is calculated as the sum of the similarities between item $i$ and those items to which user $\alpha$ is connected, normalized by the sum of the absolute values of all similarities with item $i$. 
\subsection*{Recommending ingredients to recipes}
In our study, we measure the similarity between ingredients with the aim of employing item-item collaborative filtering technique to recommend them for recipes. Using the structure of our recipes-ingredients-flavour compounds tripartite network, we outline two methodologies for computing the similarity between ingredients:
\begin{itemize}
\item Recipe-based: it evaluates ingredient similarity based on their co-occurrence in recipes. Here, the underlying hypothesis is that ingredients appearing more frequently together in recipes are more similar or compatible, reflecting traditional culinary practices and preferences. To do this we used formulas in table~\ref{tab:similarities} using matrix $R$ instead of $M$ with Greek letters denoting recipes and Latin ones ingredients;
\item Flavour-based: it assesses ingredient similarity through the shared flavor compounds between ingredients. It is grounded in the food pairing hypothesis, which posits that a greater number of common flavor compounds indicates higher compatibility within a recipe. To do this we used formulas in table~\ref{tab:similarities} using matrix $F$ instead of $M$ with Greek letters indicating flavour compounds and Latin ones ingredients
\end{itemize}
Once the similarity between ingredients has been measured we can use equation \ref{eq:density} to measure the recommendation scores of ingredients for recipes.

\subsection*{LightGCN}
LightGCN, introduced by He et al. in 2020~\cite{he2020lightgcn}, is a Graph Convolutional Network collaborative filtering. This model diverges from item-item collaborative filterings by not relying on the co-occurrence of items to determine similarity. Instead, LightGCN learns a vector representation (embedding) $e_i$ for each ingredient and $e_\alpha$ for each recipe in the dataset. The recommendation of an ingredient to a recipe \footnote{as specified and explained in the results we do not implement LightGCN for the flavor-based case} is determined by the dot product of their respective embeddings.
\begin{equation}
S_{\alpha i} = e_\alpha \cdot e_i
\label{eq:lightGCN}
\end{equation}
Owing to its machine learning foundation, LightGCN usually achieves better recommendation performance compared to item-item collaborative filterings, at the cost of exhibiting a lower degree of interpretability. However its performance depends on the value assigned to its hyperparameters. In this study, we have followed the optimization procedures detailed in the original LightGCN paper, focusing on Learning Rate, number of layers, $L_2$ regularization coefficient, and embedding size. Our optimization process identified an ideal configuration of:
\begin{itemize}
\item number of layers: 6;
\item learning rate: $10^{-2}$
\item regularization coefficient: $10^{-4}$
\item embedding size: 64
\end{itemize}
We observe that 100 epochs are sufficient for the algorithm to converge.
\subsection*{Prediction Exercise}
To assess the effectiveness of our recommendation systems, we removed a random ingredient from 10\% of the recipes and we evaluated the systems' ability to retrieve it. We denote with $\tilde{R}$ the modified biadjacency matrix of the recipe-ingredient bipartite network where the ingredients have been removed.\\
For the recipe-based approach, we followed these steps:
\begin{enumerate}
    \item We used matrix $\tilde{R}$ to measure ingredient similarity used in the methods in table \ref{tab:similarities} and to train the LightGCN model;
    \item We employed  Equation \ref{eq:density} with the $\tilde{R}$ matrix for item-item collaborative filtering and Equation \ref{eq:lightGCN} for the LightGCN model to calculate the recommendation scores for ingredients in recipes that have had one ingredient removed.
\end{enumerate}
For the flavor-based approach, the following steps were taken:
\begin{enumerate}
\item We used matrix $F$ to measure ingredient similarity based on flavor compounds;
\item We used the $\tilde{R}$ matrix and flavor-based ingredient similarity in Equation \ref{eq:density} to calculate the recommendation scores for ingredients in recipes that have had one ingredient removed.
\end{enumerate}
Once the recommendation scores were obtained, we quantified the effectiveness of the models in retrieving the removed ingredients using two performance indicators:
\begin{itemize}
\item Mean Average Precision (mAP):  the mean of the average precision \cite{salton1983introduction} across all recipes from which one ingredient have had removed. This metric communicates how highly the true ingredient is ranked in the recommendations;
\item Hit Ratio at 20 (HR@20): the fraction of recipes in which the actual removed ingredient appears among the top 20 recommendations.
\end{itemize}
To avoid dependency on the particular choice of removed ingredients, the experiment is repeated 100 times, varying the set of recipes from which an ingredient is removed.
\bibliography{sample}

\end{document}


\flushbottom
\maketitle

\thispagestyle{empty}


\section*{Score by category}
\begin{figure}[h]
    \begin{subfigure}[b]{0.49\textwidth}
        \centering
        \includegraphics[width=\linewidth]{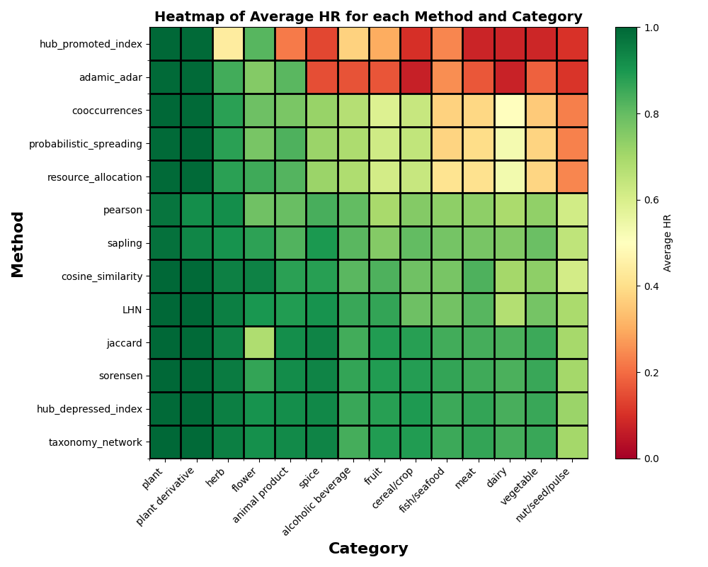}
        \subcaption{}
        \label{heatmap_cat}
    \end{subfigure}
    \begin{subfigure}[b]{0.49\textwidth}
        \centering
        \includegraphics[width=\linewidth]{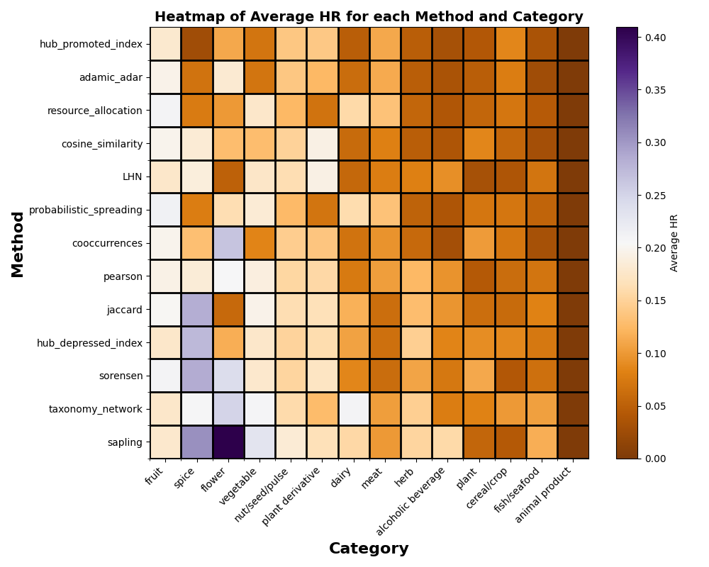}
        \subcaption{}
        \label{heatmap_cat_comp}
    \end{subfigure}
    \caption{Hit Ratio for each pairwise method recipe-based (a) and flavor-based (b) by ingredient category, computed considering the $10\%$ of the ingredients of each category for the score. For instance, 'vegetable' has $64$ ingredients, therefore the Hit Ratio grows if the removed ingredient is in the top $6$ recommendations.}
    \label{heatmap}
\end{figure}

\begin{table}[!ht]
    \centering
    \begin{tabular}{|l|l|l|}
    \hline
        \textbf{Category} & \textbf{Occurrences} & \textbf{Number of ingredients} \\ \hline
        vegetable & 97320 & 64 \\ \hline
        dairy & 63963 & 25 \\ \hline
        spice & 59977 & 27 \\ \hline
        plant derivative & 59011 & 31 \\ \hline
        cereal/crop & 41828 & 22 \\ \hline
        fruit & 34932 & 64 \\ \hline
        animal product & 28203 & 5 \\ \hline
        herb & 28126 & 23 \\ \hline
        meat & 25498 & 24 \\ \hline
        nut/seed/pulse & 14881 & 20 \\ \hline
        alcoholic beverage & 9334 & 25 \\ \hline
        fish/seafood & 7899 & 27 \\ \hline
        plant & 5071 & 14 \\ \hline
        flower & 213 & 12 \\ \hline
    \end{tabular}
    \caption{Occurrences in recipes and number of ingredients for each category. There is no obvious relationship between these values and the score by category, confirming pairwise methods are not able to totally grasp the context in which ingredients are used.}
\end{table}

\section*{Score by ingredient}
\begin{figure}[h]
    \begin{subfigure}[b]{0.49\textwidth}
        \centering
        \includegraphics[width=\linewidth]{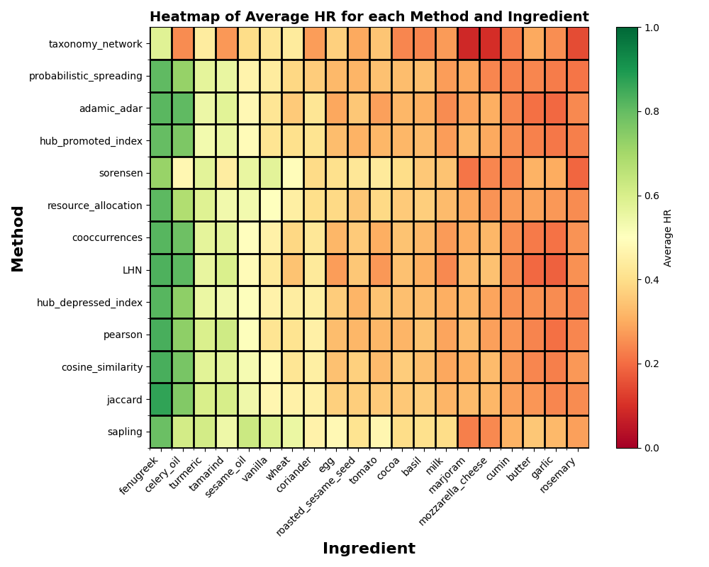}
        \subcaption{}
        \label{heatmap_ing}
    \end{subfigure}
    \begin{subfigure}[b]{0.49\textwidth}
        \centering
        \includegraphics[width=\linewidth]{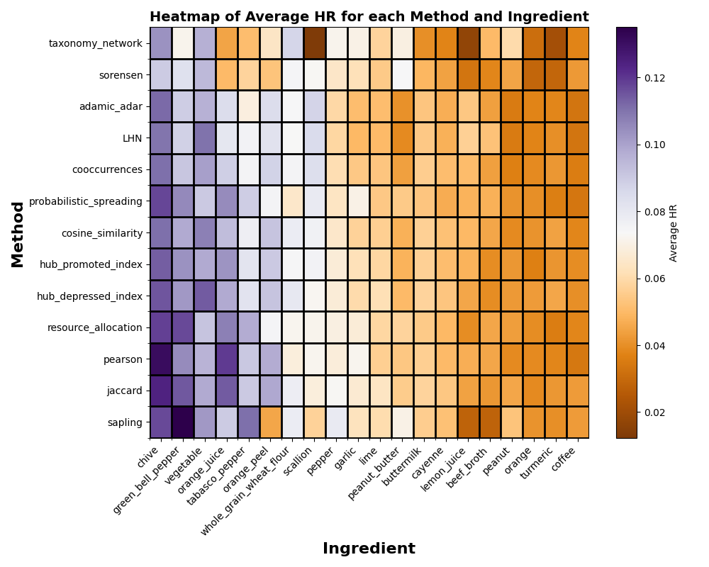}
        \subcaption{}
        \label{heatmap_ing_comp}
    \end{subfigure}
    \caption{Hit Ratio for each pairwise method recipe-based (a) and flavor-based (b) by single ingredient, computed like in the previous case. Only the ingredients with the best score are reported, confirming this time the number of occurrences significantly relates with the performance for that ingredient (as we expect).}
    \label{heatmap}
\end{figure}

\begin{table}[!ht]
    \centering
    \begin{tabular}{|l|l|l|l|}
    \hline
        \textbf{Ingredient} & \textbf{Average \# of removals} & \textbf{Occurrences} & \textbf{Category} \\ \hline
        chive & 106 & 1337 & vegetable \\ \hline
        green\_bell\_pepper & 186 & 2590 & vegetable \\ \hline
        vegetable & 141 & 1705 & vegetable \\ \hline
        orange\_juice & 200 & 1728 & fruit \\ \hline
        tabasco\_pepper & 75 & 977 & spice \\ \hline
        orange\_peel & 64 & 596 & plant \\ \hline
        whole\_grain\_wheat\_flour & 62 & 734 & cereal/crop \\ \hline
        scallion & 391 & 4798 & vegetable \\ \hline
        pepper & 820 & 9282 & spice \\ \hline
        garlic & 1474 & 17465 & vegetable \\ \hline
        lime & 135 & 1165 & fruit \\ \hline
        peanut\_butter & 134 & 1029 & plant derivative \\ \hline
        buttermilk & 185 & 1637 & dairy \\ \hline
        cayenne & 745 & 8303 & spice \\ \hline
        lemon\_juice & 511 & 5099 & fruit \\ \hline
        beef\_broth & 61 & 850 & meat \\ \hline
        peanut & 73 & 512 & nut/seed/pulse \\ \hline
        orange & 228 & 1728 & fruit \\ \hline
        turmeric & 95 & 1329 & spice \\ \hline
        cofeee & 98 & 720 & plant derivative \\ \hline
    \end{tabular}
\end{table}

\bibliography{sample}